\documentstyle[11pt,newpasp,epsf]{article}

\begin{document}

\title{Abundance analysis of planetary host stars}

\author{U. Heiter and R. E. Luck}
\affil{Department of Astronomy, Case Western Reserve University, Cleveland, OH 44106}





\begin{abstract}
We present atmospheric parameters and Fe abundances derived for the majority of dwarf stars (north of -30 degrees declination) which are up to now known to host extrasolar planets. High-resolution spectra have been obtained with the Sandiford Echelle spectrograph on the 2.1m telescope at the University of Texas McDonald Observatory. We have used the same model atmospheres, atomic data and equivalent width modeling program for the analysis of all stars. Abundances have been derived differentially to the Sun, using a solar spectrum obtained with Callisto as the reflector with the same instrumentation. A similar analysis has been performed for a sample of stars for which radial velocity data exclude the presence of a close-in giant planetary companion. The results are compared to the recent studies found in the literature.
\end{abstract}


\keywords{abundances,atmospheric parameters,solar-type stars,extrasolar planets}


%
%
%

\section{Introduction}
\begin{figure}
\plotfiddle{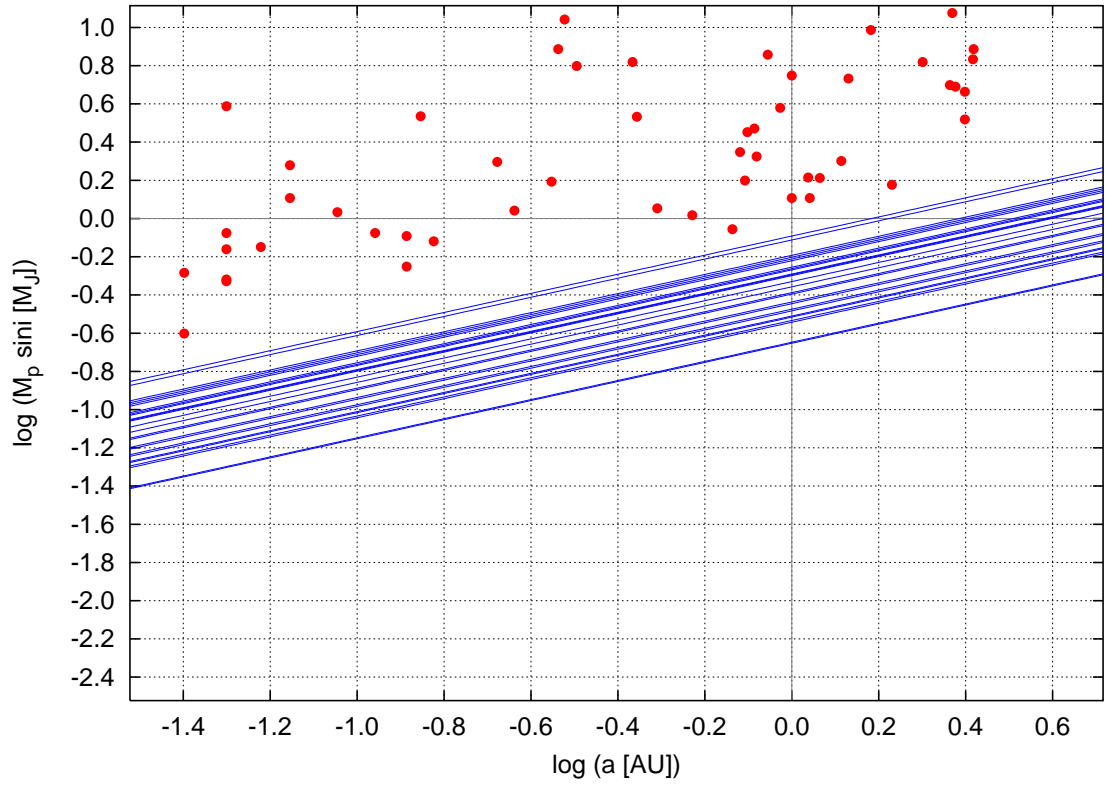}{11cm}{270}{60}{60}{-220}{350}
\caption{Red dots: Masses and semi-major axes of currently known extrasolar planets. 
Blue lines: Upper (mass) and lower (orbital radius) limits for stars from the Lick survey (Cumming et al.1999).}
\label{limits}
\end{figure}

To examine the relation between the metallicity of the stellar host and the existence of close-in massive planets, we have analyzed high-resolution, high signal-to-noise spectra of the dwarf stars with super-Jupiter planets listed in the ``Extrasolar Planets Encyclopaedia''\footnote{http://www.obspm.fr/encycl/encycl.html} (44 stars at the time when this poster was prepared, hereafter CGP dwarfs). Note that this list does not include stars with companions with $M_{\rm p} \sin i > 12 M_{\rm J}$. A list of comparison stars was compiled using the results of the Lick planet search (Cumming et al. 1999). We analyzed all stars for which radial velocity data exclude the presence of a close-in giant planetary companion (23 stars, hereafter no-CGP dwarfs). The selection process is illustrated in Figure~\ref{limits}, which shows the upper limits for planetary masses (in Jupiter-masses) as a function of orbital radii (in AU) for this sample, as derived from radial-velocity data, and masses and semi-major axes for known extrasolar planets. Those of the latter which would fall below the lower limit for the comparison sample have been excluded. 

\section{Observations}
The spectra were obtained at the 2.1m telescope of the McDonald Observatory with the Sandiford Echelle spectrograph. They cover a wavelength range from about 4840 to 7000~\AA\ with a resolution of 60000. The list of the observed stars can be found in Tables~\ref{dwp_obs} and \ref{dwop_obs}. The reductions were done with IRAF (echelle order extraction) and a Windows based graphical package (ASP) for continuum and wavelength setting and equivalent width determination developed by REL. A comparison of equivalent widths with published values indicates agreement at the 5\% level.

\begin{table}
\caption{Observations of CGP dwarfs.}
\label{dwp_obs}
\begin{center}
\begin{tabular}{rc}
HD & date \\
\tableline\\[-3mm]
8574        &  2001-08-23  \\
            &  2001-08-27  \\
9826        &  1999-10-19  \\
            &  1999-10-23  \\
10697       &  1992-11-03  \\
            &  1992-11-05  \\
            &  1992-11-07  \\
            &  1997-02-03  \\
            &  1999-10-19  \\
12661       &  2000-08-18  \\
            &  2000-08-21  \\
19994       &  2000-10-20  \\
            &  2001-02-11  \\
            &  2001-08-23  \\
            &  2001-08-25  \\
28185       &  2001-10-25  \\
            &  2001-10-27  \\
33636       &  2002-01-23  \\
            &  2002-01-26  \\
37124       &  2000-01-26  \\
            &  2000-01-29  \\
38529       &  1992-11-04  \\
            &  1992-11-05  \\
            &  1992-11-07  \\
            &  1993-03-10  \\
            &  1997-01-31  \\
            &  1997-02-02  \\
50554       &  2001-10-23  \\
            &  2001-10-27  \\
52265       &  2001-10-24  \\
            &  2001-10-27  \\
68988       &  2002-01-24  \\
            &  2002-01-26  \\
74156       &  2001-10-25  \\
            &  2001-10-28  \\
75732       &  1993-03-03  \\
            &  1993-03-05  \\
            &  1993-03-11  \\
            &  1993-03-07  \\
\tableline
\end{tabular}
\begin{tabular}{rc}
HD & date \\
\tableline\\[-3mm]
            &  1997-02-03  \\
            &  1999-10-19  \\
80606       &  2002-01-23  \\
            &  2002-01-28  \\
82943       &  2002-01-22  \\
            &  2002-01-25  \\
92788       &  2001-05-08  \\
            &  2001-05-14  \\
89744       &  2001-05-08  \\
            &  2001-05-14  \\
106252      &  2001-05-08  \\
            &  2001-05-14  \\
114762      &  1999-05-26  \\
            &  1999-05-28  \\
            &  2000-01-25  \\
117176      &  1999-05-25  \\
            &  1999-05-28  \\
            &  2000-01-25  \\
120136      &  1999-05-25  \\
            &  1999-05-28  \\
            &  2000-01-25  \\
130322      &  2000-01-26  \\
            &  2000-01-30  \\
134987      &  2000-04-25  \\
            &  2000-04-30  \\
136118      &  2002-05-21  \\
            &  2002-05-24  \\
141937      &  2001-05-08  \\
            &  2002-01-26  \\
143761      &  1999-05-26  \\
            &  1999-05-28  \\
            &  2001-05-10  \\
145675      &  1993-03-06  \\
            &  1993-03-09  \\
            &  1993-03-11  \\
            &  1997-04-24  \\
            &  1998-05-15  \\
168443      &  1999-05-30  \\
            &  1999-10-25  \\
\tableline
\end{tabular}
\begin{tabular}{rc}
HD & date \\
\tableline\\[-3mm]
169830      &  2000-08-18  \\
            &  2000-08-21  \\
177830      &  2000-08-18  \\
            &  2000-08-21  \\
178911      &  2001-05-08  \\
            &  2001-05-14  \\
            &  2001-08-27  \\
179949      &  2001-08-24  \\
            &  2001-08-28  \\
186427      &  1999-05-26  \\
            &  1999-05-28  \\
            &  2000-08-19  \\
187123      &  1999-05-30  \\
            &  1999-10-25  \\
190228      &  2000-10-22  \\
            &  2001-08-23  \\
            &  2001-08-27  \\
192263      &  2000-08-19  \\
            &  2000-08-21  \\
195019      &  1999-05-30  \\
            &  1999-10-25  \\
209458      &  2000-08-18  \\
            &  2000-08-21  \\
210277      &  1999-10-19  \\
            &  1999-10-25  \\
217014      &  1992-11-03  \\
            &  1992-11-06  \\
            &  1992-11-07  \\
            &  1997-10-18  \\
            &  1999-10-19  \\
217107      &  1999-10-19  \\
            &  1999-10-24  \\
222582      &  2000-08-18  \\
            &  2000-08-21  \\
BD -10 3166 &  2001-05-10  \\
            &  2002-01-26  \\
            &  2002-01-28  \\
Callisto    &  2001-10-26  \\
            &  2001-10-28  \\
\tableline
\end{tabular}
\end{center}
\end{table}

\begin{table}
\caption{Observations of no-CGP dwarfs.}
\label{dwop_obs}
\begin{center}
\begin{tabular}{rc}
HD & date \\
\tableline\\[-3mm]
166    & 1999-10-25 \\
       & 2000-08-22 \\
4628   & 1999-10-25 \\
       & 2000-08-20 \\
       & 2000-08-22 \\
10476  & 2000-08-20 \\
       & 2001-08-23 \\
12235  & 1992-11-04 \\
       & 1992-11-06 \\
       & 1992-11-08 \\
       & 1992-11-09 \\
       & 1997-02-03 \\
       & 1999-10-19 \\
16160  & 2000-08-18 \\
       & 2001-08-25 \\
16895  & 2000-08-19 \\
       & 2001-10-23 \\
22484  & 1999-10-21 \\
       & 1999-10-24 \\
26965  & 1999-10-19 \\
       & 1999-10-23 \\
       & \\
       & \\
       & \\
\tableline
\end{tabular}
\begin{tabular}{rc}
HD & date \\
\tableline\\[-3mm]
32147  & 1992-11-03 \\
       & 1992-11-05 \\
       & 1992-11-08 \\
       & 1993-03-10 \\
       & 1995-10-11 \\
       & 1997-01-31 \\
       & 1997-02-02 \\
48682  & 2000-10-20 \\
       & 2001-10-25 \\
       & 2001-10-27 \\
50281  & 2001-10-26 \\
       & 2002-01-25 \\
76151  & 1993-03-04 \\
       & 1993-03-06 \\
       & 1993-03-07 \\
       & 1993-03-10 \\
       & 1997-02-03 \\
       & 2000-01-29 \\
84737  & 2002-05-21 \\
       & 2002-05-24 \\
126053 & 2002-05-21 \\
       & 2002-05-24 \\
       & \\
       & \\
\tableline
\end{tabular}
\begin{tabular}{rc}
HD & date \\
\tableline\\[-3mm]
149661 & 1993-03-03 \\
       & 1993-03-05 \\
       & 1993-03-07 \\
       & 1993-03-10 \\
       & 1997-04-24 \\
       & 1997-04-28 \\
157214 & 2000-04-25 \\
       & 2000-04-29 \\
170657 & 2001-08-24 \\
       & 2001-08-27 \\
166620 & 2000-10-19 \\
       & 2001-08-25 \\
       & 2001-08-27 \\
185144 & 1999-10-19 \\
       & 1999-10-23 \\
186408 & 1999-05-26 \\
       & 1999-05-28 \\
       & 2000-08-19 \\
201091 & 1997-08-17 \\
       & 1999-10-19 \\
219134 & 1999-10-19 \\
       & 1999-10-24 \\
222368 & 2000-08-19 \\
       & 2000-08-21 \\
\tableline
\end{tabular}
\end{center}
\end{table}

\section{Analysis}
\begin{figure}
\plotfiddle{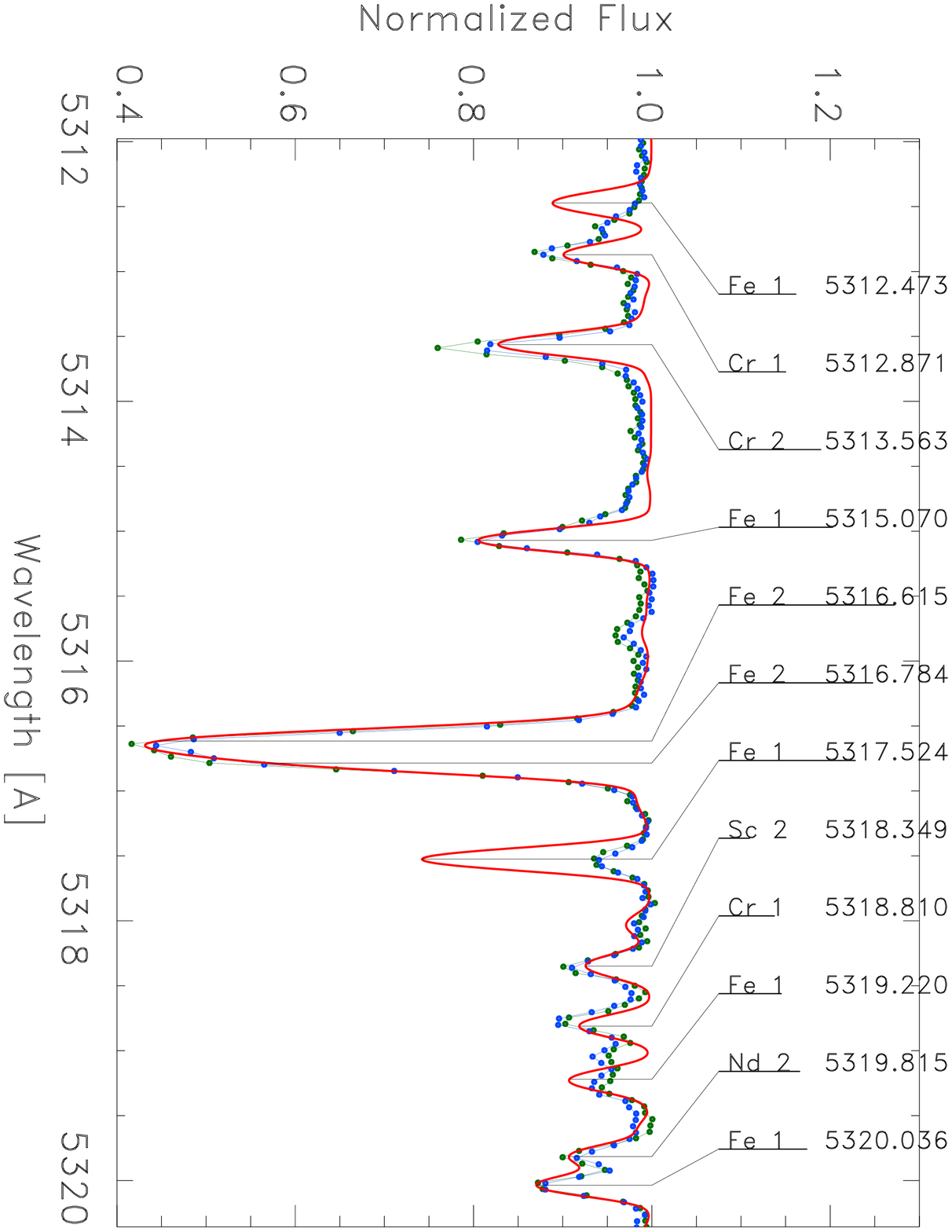}{11cm}{90}{60}{60}{250}{0}
\caption{Sample spectra of HD 195019 (green dots) and the Sun (Callisto, blue dots). 
Red line: synthetic spectrum. Note that for the Fe~I line at 5317~\AA\ the $gf$ value is obviously wrong, but this source of error is canceled out in a strictly differential analysis.}
\label{spectra}
\end{figure}

The abundance analysis was done strictly differentially with respect to the Sun using a large sample of lines (cf. Figures~\ref{trends1}a,b). We have obtained a solar flux spectrum (Callisto) using the same instrumentation and reduction procedure as used for our program stars. Figure~\ref{spectra} shows a portion of the observed spectra of HD 195019 and the Sun, as well as a synthetic spectrum, computed with SYNTH (Piskunov 1992).
Synthetic equivalent widths were calculated and fit to the observations by variation of abundance for each line with the program LINES (originally by Sneden 1974). We used MARCS model atmospheres (Gustafsson et al. 1975) and an atomic data list compiled by REL. The relative abundance for each line was then calculated by subtracting the abundance obtained from the Callisto spectrum for the corresponding line. From all Fe lines available within the observed spectral range, only those giving an abundance within two standard deviations (2$\sigma$) from the mean have been retained for the derivation of the atmospheric parameters described in the following subsection. This was done in order to discard ``outliers'' automatically. For the calculation of the final mean abundances, however, the line list has been further reduced by rejecting all lines with an abundance deviation from the mean of more than 1$\sigma$, which decreased the errors of the abundances significantly.

\subsection{Atmospheric parameters}
\begin{figure}
\plotfiddle{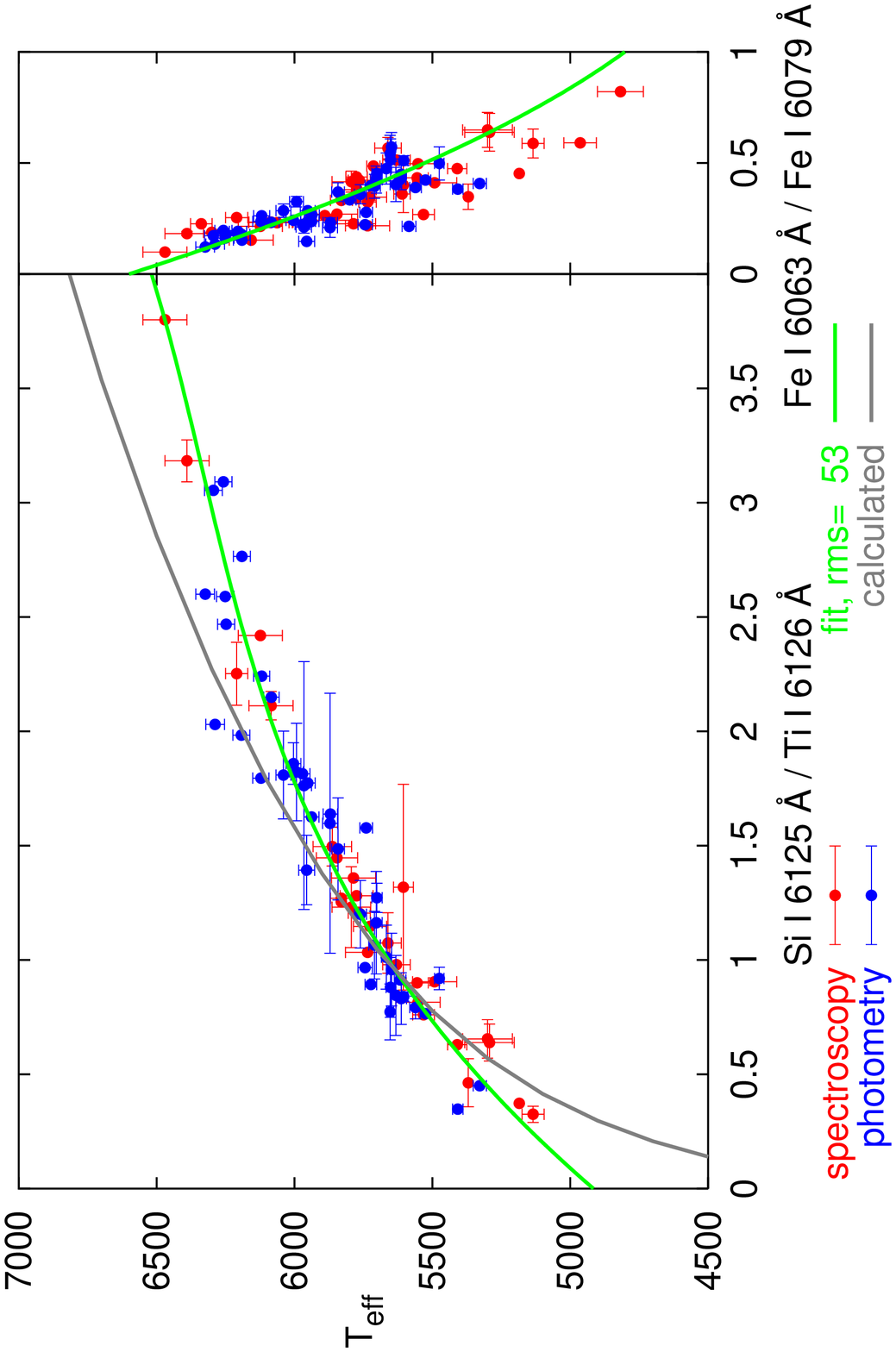}{11cm}{270}{60}{60}{-220}{350}
\caption{Two examples for the dependence of line-depth ratios on effective temperature. Red: literature data, blue: Geneva photometry, green: polynomial fit, gray: relation for ratios calculated from synthetic spectra. The relation shown in the right panel has not been used for the calibration.}
\label{ratios}
\end{figure}

To determine the atmospheric parameters, iron line abundances were calculated
for each star for a small ($T_{\rm eff}$, $\log g$, $v_{\rm micro}$) -- grid, centered on $v_{\rm micro}$ = 1~km\,s$^{-1}$ and ($T_{\rm eff}$, $\log g$) determined by one of three methods in the following order of preference:
\begin{enumerate}
\item Literature data (previous abundance analyses\footnote{Santos et al. (2000, 2001), Gonzalez et al. (1997--2001), Fuhrmann (1998), Edvardsson et al. (1993)})
\item Calibration for Geneva photometry (K\"unzli et al. 1997)
\item A new calibration for line-depth ratios and $T_{\rm eff}$ for our sample of dwarf stars. We used 20 of the 32 line-pairs which Kovtyukh and Gorlova (2000) had used for a similar calibration for supergiants, our observations, and effective temperatures from methods 1 and 2. An example for a useful (left panel) and a rejected (right panel) line-pair is shown in Figure~\ref{ratios}. The line-ratio -- $T_{\rm eff}$ relations were obtained from polynomial fits to the data, or calculated from synthetic spectra in some cases where only few datapoints were available.
\end{enumerate}

\begin{figure}
\plotfiddle{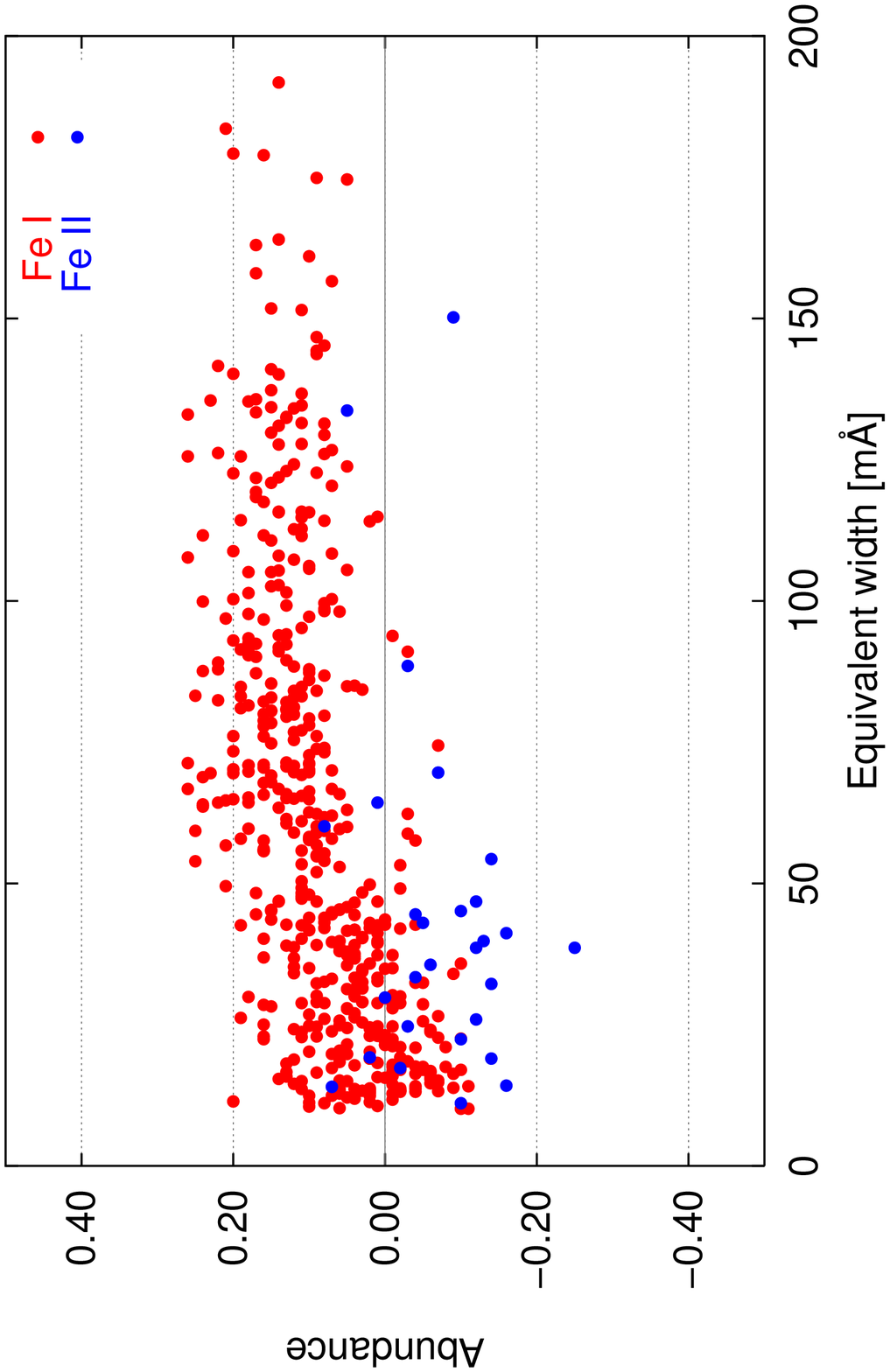}{9cm}{270}{50}{50}{-220}{300}
\plotfiddle{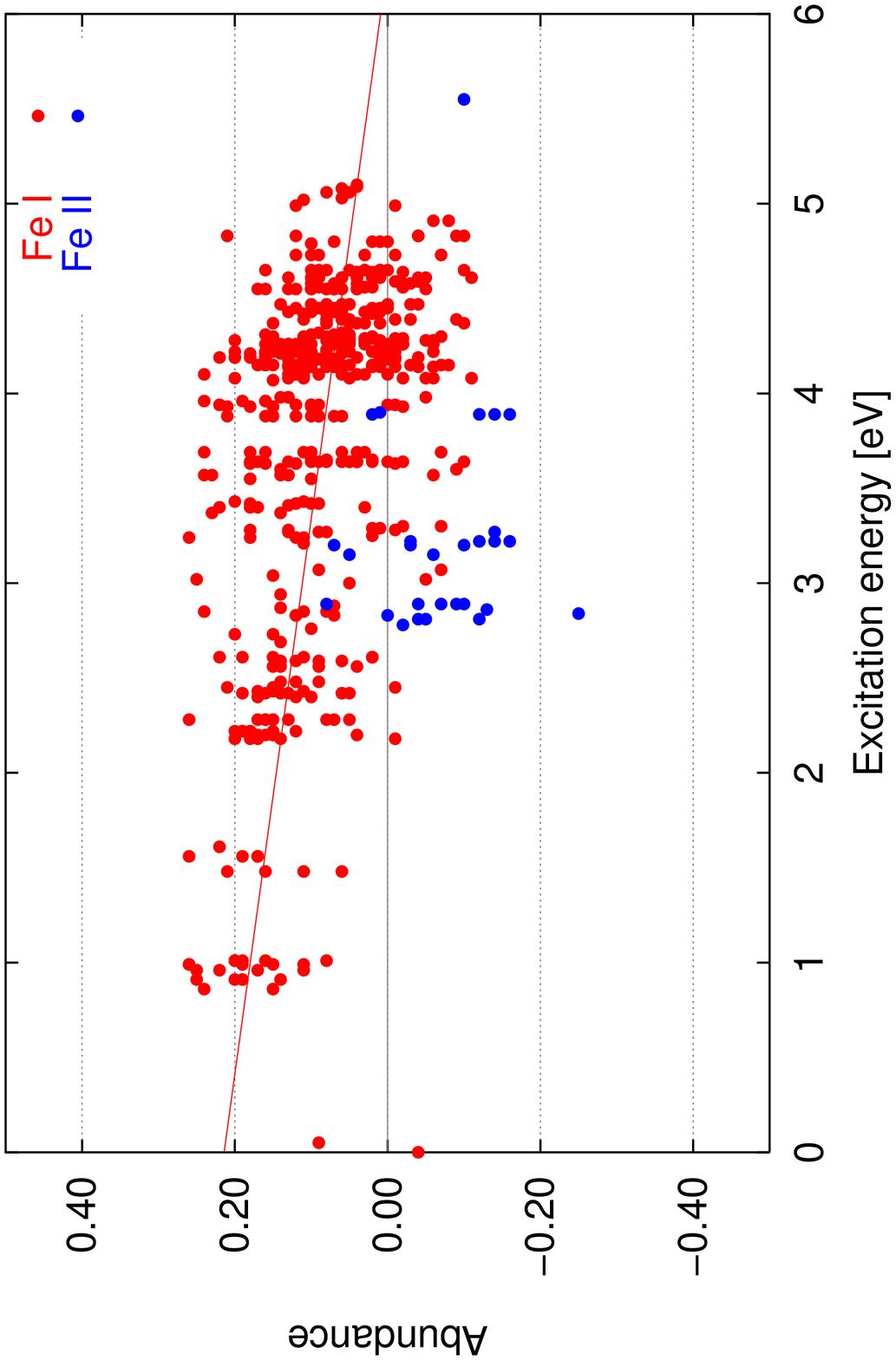}{9cm}{270}{50}{50}{-220}{300}
\caption{Fe abundances in dex (0.00\dots[Fe]$_{\sun}$) versus atomic data for HR 5072 and ($T_{\rm eff}$, $\log g$, $v_{\rm micro}$) = (5700, 4.0, 0.7). These parameters obviously do not correspond to the best-fit values.}
\label{trends1}
\end{figure}

\begin{figure}
\plotfiddle{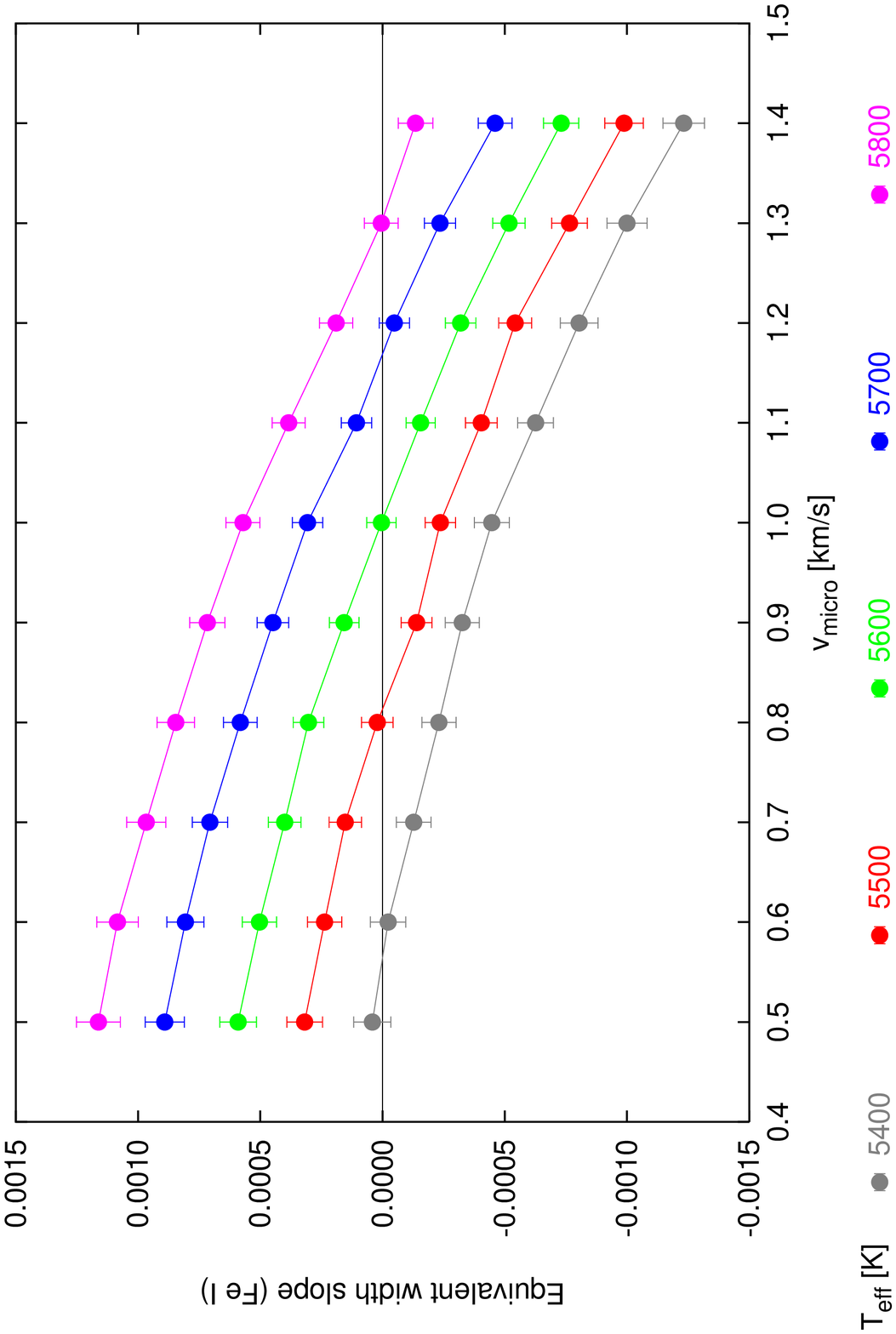}{9cm}{270}{50}{50}{-220}{300}
\plotfiddle{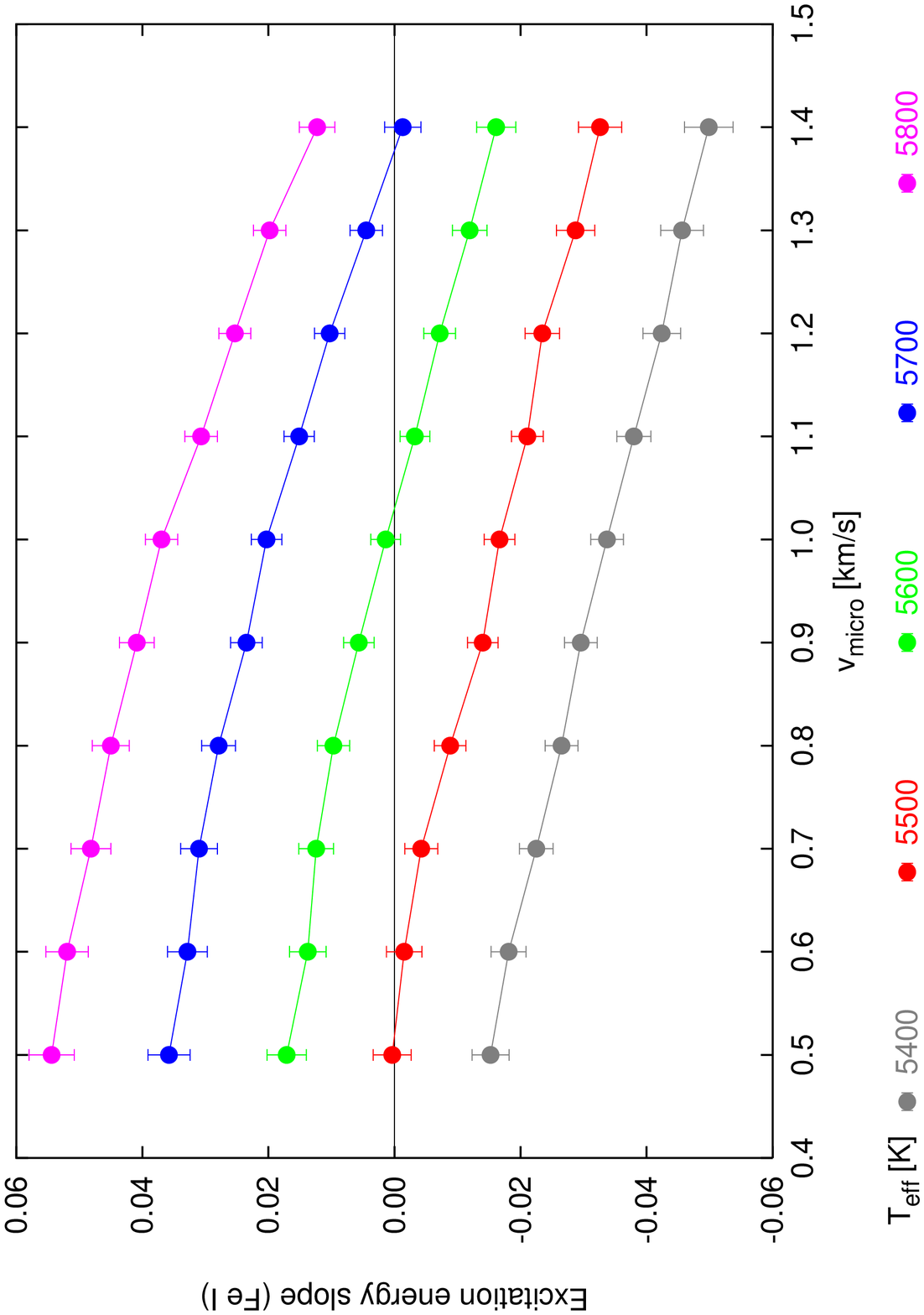}{9cm}{270}{50}{50}{-220}{300}
\caption{Dependence of the slopes appearing in diagrams like those shown in Figure~\ref{trends1} on $v_{\rm micro}$ and $T_{\rm eff}$, for Fe~I, here for $\log g$=4.1 (HR 5072). The errorbars are determined from linear regression.}
\label{trends3}
\end{figure}

\begin{figure}
\plotfiddle{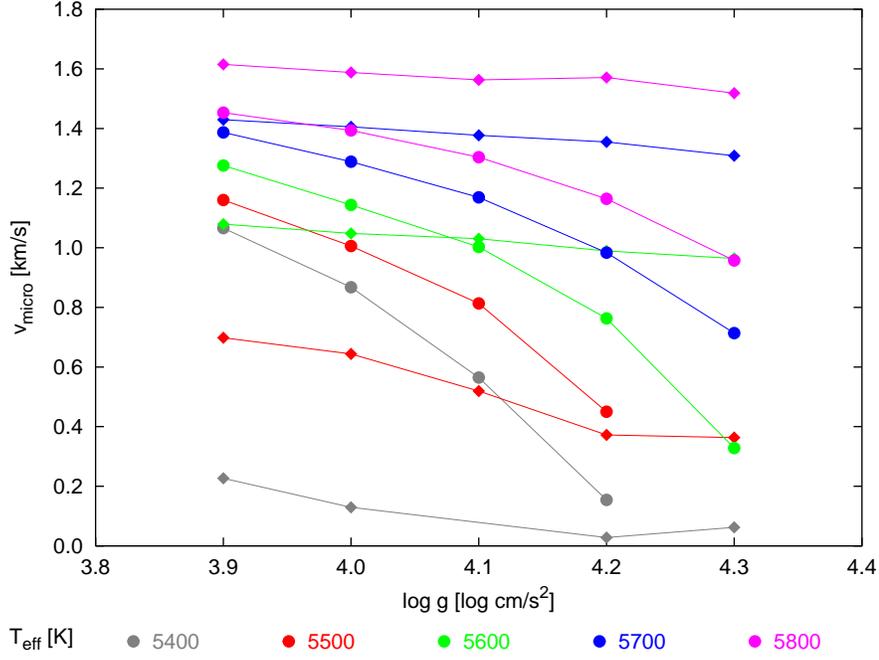}{8cm}{270}{50}{50}{-220}{270}
\caption{Values of $v_{\rm micro}$, for which the slopes in the abundance versus atomic data diagrams vanish, in dependence of $T_{\rm eff}$ and $\log g$. Dots and diamonds correspond to equivalent width and excitation energy, respectively.}
\label{trends5}
\end{figure}

\begin{figure}
\plotfiddle{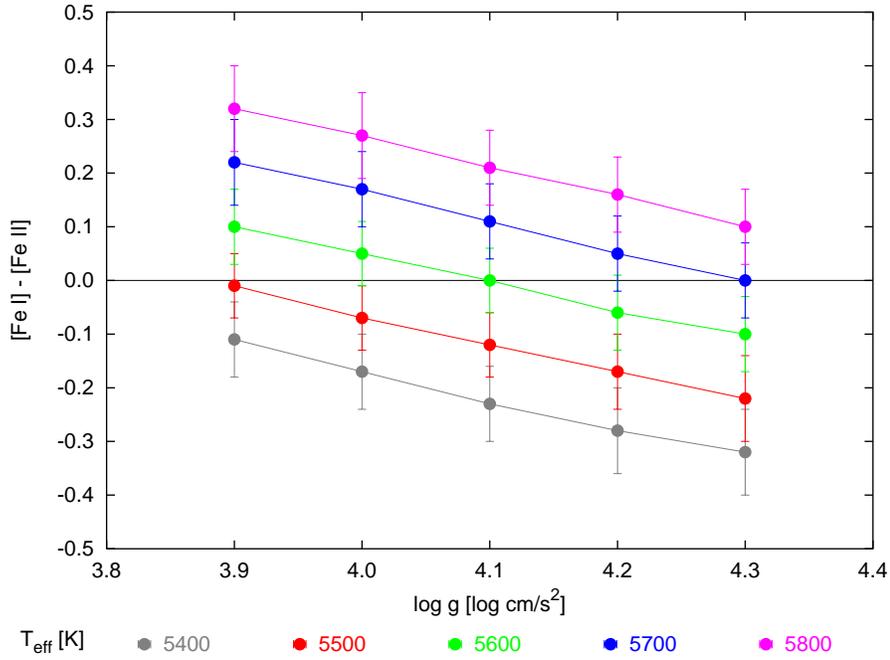}{8cm}{270}{50}{50}{-220}{270}
\caption{Difference between the mean abundances of neutral and ionized iron, for $v_{\rm micro}$=1.0~km\,s$^{-1}$.The errorbars correspond to the standard deviation of the line abundances.}
\label{trends6}
\end{figure}

\begin{figure}
\plotfiddle{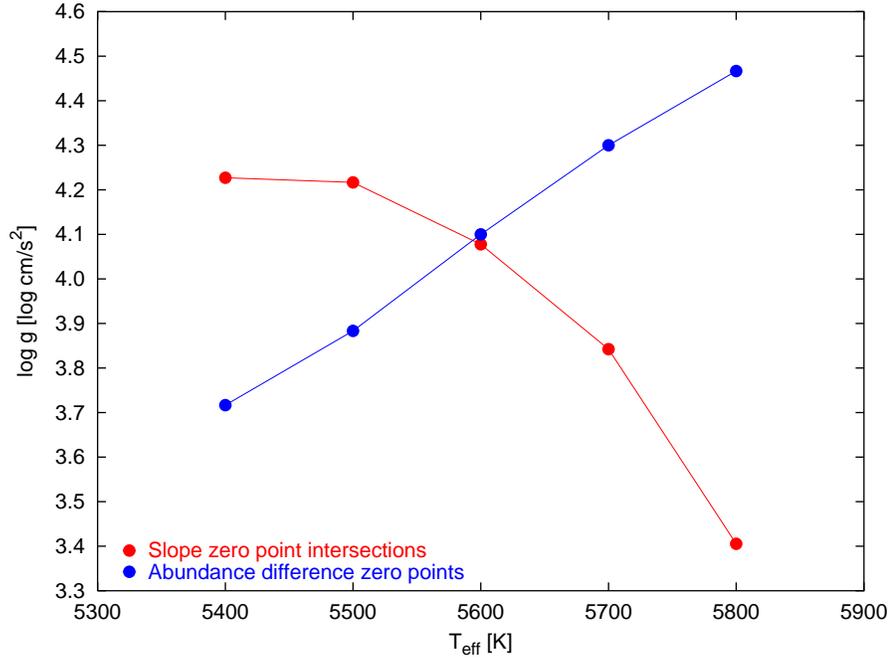}{8cm}{270}{50}{50}{-220}{270}
\caption{Red: Intersections of two lines from Figure~\ref{trends5} corresponding to equivalent width and excitation energy. Blue: Combinations of $T_{\rm eff}$ and $\log g$, for which the difference between the mean abundances of neutral and ionized iron is zero (determined from Figure~\ref{trends6}).}
\label{trends7}
\end{figure}

\begin{figure}
\plotfiddle{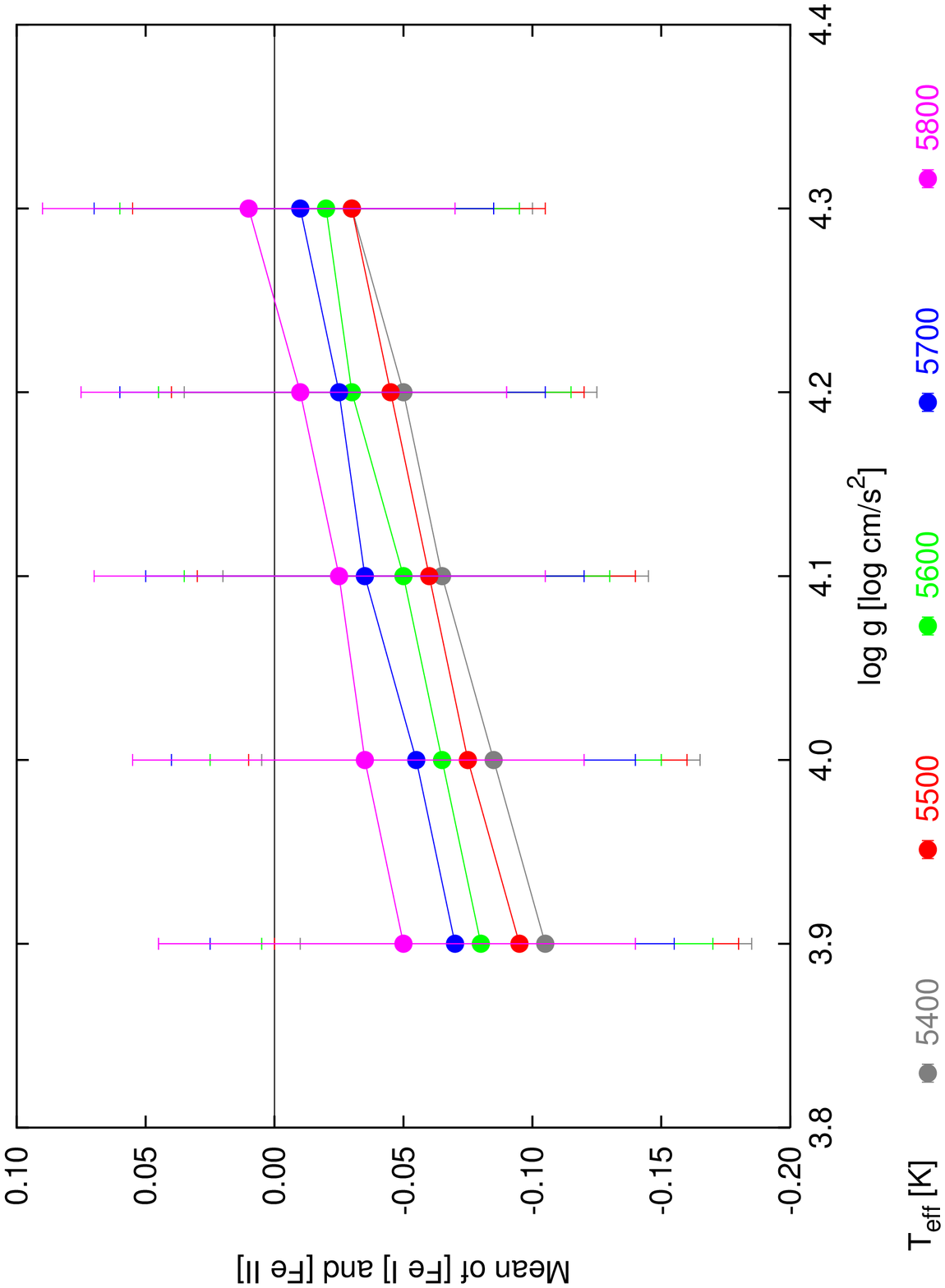}{8cm}{270}{50}{50}{-220}{270}
\caption{Variation of the mean iron abundance with parameters. Here, the error bars indicate the abundance range covered by using the $v_{\rm micro}$ range of Figure~\ref{trends3}.}
\label{trends8}
\end{figure}

The final atmospheric parameters were determined by driving the slopes of the line-strength and excitation energy relations with Fe~I line abundance to zero and demanding ionization equilibrium. As an example, Figure~\ref{trends1} shows the trends visible in an abundance versus equivalent width or excitation energy diagram, if $T_{\rm eff}$, $\log g$, and $v_{\rm micro}$ deviate from the best-fit values by 100~K, 0.1~[cgs], and 0.3~km\,s$^{-1}$, respectively (for the planet host HR 5072). We determine the best-fit values by first identifying the value of $v_{\rm micro}$, for which the trends in the above mentioned diagrams are zero, for each model in the ($T_{\rm eff}$, $\log g$) grid (cf. Figure~\ref{trends3}). Figure~\ref{trends5} shows these ``zero-points'' for equivalent width (dots) and excitation energy (diamonds) for each grid point. From the set of ($T_{\rm eff}$, $\log g$) models for which the two zero-points are equal (i.e. the intersections of two corresponding lines in Figure~\ref{trends5}), we can finally identify one parameter pair, which gives the same abundance for neutral and ionized Fe lines. Figure~\ref{trends6} illustrates the dependence of the abundance difference [Fe~I]$-$[Fe~II] on $T_{\rm eff}$ and $\log g$. The dependence of this quantity on $v_{\rm micro}$ is negligible. The combination of information extracted from Figures~\ref{trends5} and \ref{trends6} is shown in Figure~\ref{trends7}. Figure~\ref{trends8} shows the variation of the mean Fe abundance within the parameter grid. From these diagrams we estimate the errors in our final spectroscopically determined parameters to be 50 K, 0.05 [cgs] and 0.05 km\,s$^{-1}$, for $T_{\rm eff}$, $\log g$, and $v_{\rm micro}$, respectively.

\section{Results}

\subsection{Atmospheric parameters}
A comparison of the parameters derived as described above with parameters found in the literature and derived from photometry results in the mean differences and standard deviations listed in Table~\ref{parameters} (parameter[this work] $-$ parameter[comparison]).

\begin{table}
\caption{Differences between parameters determined in this work, in the literature and with photometric calibrations.}
\label{parameters}
\begin{center}
\begin{tabular}{lrr@{}rrr}
                                          &  N & \multicolumn{2}{c}{$T_{\rm eff}$} & \multicolumn{1}{c}{$\log g$} & \multicolumn{1}{c}{[Fe]} \\
\tableline\\[-3mm]
Spectroscopy\tablenotemark{1}             & 52\tablenotemark{\ast} &  55 $\pm$ &  90 & 0.05 $\pm$ 0.20 &  0.02 $\pm$ 0.07 \\
\qquad CGP dwarfs                         & 37 &  50 $\pm$ &  90 & 0.05 $\pm$ 0.20 &  0.02 $\pm$ 0.08 \\
\qquad No-CGP dwarfs                      &  9 &  80 $\pm$ & 120 & 0.10 $\pm$ 0.20 &  0.02 $\pm$ 0.07 \\
Geneva photometry\tablenotemark{2,3}      & 41\tablenotemark{\ast}  &  45 $\pm$ &  75 & 0.00 $\pm$ 0.35 &  0.10 $\pm$ 0.09 \\
Str\"omgren photometry\tablenotemark{3,4} & 50\tablenotemark{\ast}  & 190 $\pm$ & 110 & 0.25 $\pm$ 0.70 & $-$0.10 $\pm$ 0.66 \\
\tableline
\end{tabular}
\tablenotetext{1}{Santos et al. (2000, 2001), Gonzalez et al. (1997--2001), Fuhrmann (1998), Edvardsson et al. (1993)}
\tablenotetext{2}{Calibration by K\"unzli et al. (1997)}
\tablenotetext{3}{Data from the GCPD (Mermilliod et al. 1997)}
\tablenotetext{4}{Calibration by Napiwotzki et al. (1993)}
\tablenotetext{\ast}{The following solar type stars are included in addition to that listed in Tables~\ref{results1} and \ref{results2}: HD\,16141, HD\,18445, HD\,22049, HD\,29587, HD\,95128, HD\,110833, HD\,112758, HD\,140913, HD\,168746, HD\,178911, HD\,202206.}
\end{center}
\end{table}

\begin{table}
\caption{Parameters and Fe abundances for CGP dwarfs.}
\label{results1}
\begin{center}
\begin{tabular}{rrrrr@{$\pm$}r}
\multicolumn{1}{c}{HD} & \multicolumn{1}{c}{$T_{\rm eff}$} & \multicolumn{1}{c}{$\log g$} & \multicolumn{1}{c}{$v_{\rm micro}$} & \multicolumn{1}{c}{[Fe]} &  \\
\tableline\\[-3mm]
8574 & 6100 & 4.30 & 1.45 & $-$0.04 & 0.05 \\
9826 & 6200 & 4.40 & 1.60 & 0.13 & 0.07 \\
10697 & 5650 & 4.05 & 0.90 & 0.16 & 0.06 \\
12661 & 5700 & 4.10 & 0.50 & 0.39 & 0.07 \\
19994 & 6150 & 4.30 & 1.50 & 0.23 & 0.07 \\
28185 & 5700 & 4.20 & 0.80 & 0.24 & 0.06 \\
33636 & 6050 & 4.55 & 1.25 & $-$0.11 & 0.05 \\
37124 & 5700 & 4.50 & 1.10 & $-$0.38 & 0.05 \\
38529 & 5750 & 4.05 & 1.50 & 0.48 & 0.12 \\
50554 & 6050 & 4.50 & 1.20 & $-$0.04 & 0.04 \\
52265 & 6100 & 4.40 & 1.30 & 0.15 & 0.05 \\
68988 & 6000 & 4.45 & 1.35 & 0.36 & 0.06 \\
74156 & 6050 & 4.35 & 1.25 & 0.09 & 0.05 \\
75732 & 5500 & 4.40 & 1.00 & 0.55 & 0.16 \\
80606 & 5700 & 4.40 & 0.90 & 0.46 & 0.07 \\
82943 & 5900 & 4.40 & 0.75 & 0.24 & 0.04 \\
89744 & 6300 & 4.40 & 1.80 & 0.22 & 0.08 \\
92788 & 5700 & 4.20 & 0.50 & 0.30 & 0.06 \\
106252 & 5890 & 4.40 & 1.10 & $-$0.10 & 0.04 \\
114762 & 6000 & 4.40 & 1.50 & $-$0.78 & 0.06 \\
117176 & 5600 & 4.10 & 1.00 & $-$0.05 & 0.03 \\
120136 & 6600 & 4.70 & 1.90 & 0.37 & 0.12 \\
130322 & 5400 & 4.40 & 0.00 & 0.10 & 0.06 \\
134987 & 5720 & 4.25 & 0.70 & 0.35 & 0.11 \\
136118 & 6250 & 4.45 & 1.60 & $-$0.03 & 0.07 \\
141937 & 5900 & 4.45 & 0.90 & 0.12 & 0.04 \\
143761 & 5900 & 4.40 & 1.25 & $-$0.25 & 0.04 \\
145675 & 5500 & 4.30 & 1.10 & 0.59 & 0.15 \\
168443 & 5600 & 4.10 & 0.80 & 0.09 & 0.04 \\
169830 & 6300 & 4.40 & 1.60 & 0.09 & 0.06 \\
177830 & 5000 & 3.70 & 0.90 & 0.61 & 0.14 \\
178911 & 6050 & 4.50 & 1.10 & 0.15 & 0.06 \\
179949 & 6200 & 4.50 & 1.20 & 0.20 & 0.06 \\
186427 & 5800 & 4.40 & 0.95 & 0.06 & 0.03 \\
187123 & 5800 & 4.35 & 0.90 & 0.08 & 0.04 \\
190228 & 5360 & 3.90 & 0.90 & $-$0.17 & 0.04 \\
192263 & 5100 & 4.45 & 0.00 & 0.11 & 0.10 \\
195019 & 5830 & 4.30 & 1.05 & 0.05 & 0.04 \\
209458 & 6100 & 4.50 & 1.30 & $-$0.02 & 0.05 \\
210277 & 5700 & 4.40 & 1.20 & 0.27 & 0.06 \\
217014 & 5750 & 4.25 & 0.70 & 0.24 & 0.05 \\
217107 & 5750 & 4.35 & 1.15 & 0.39 & 0.06 \\
222582 & 5800 & 4.40 & 0.80 & 0.03 & 0.04 \\
BD $-$10 3166 & 5550 & 4.45 & 1.00 & 0.51 & 0.10 \\
\tableline
\end{tabular}
\end{center}
\end{table}

\begin{table}
\caption{Parameters and Fe abundances for no-CGP dwarfs.}
\label{results2}
\begin{center}
\begin{tabular}{rrrrr@{$\pm$}r}
\multicolumn{1}{c}{HD} & \multicolumn{1}{c}{$T_{\rm eff}$} & \multicolumn{1}{c}{$\log g$} & \multicolumn{1}{c}{$v_{\rm micro}$} & \multicolumn{1}{c}{[Fe]} & \\
\tableline\\[-3mm]
166 & 5550 & 4.50 & 0.80 & 0.13 & 0.05 \\
4628 & 5150 & 4.60 & 0.80 & $-$0.21 & 0.08 \\
10476 & 5200 & 4.35 & 0.00 & 0.03 & 0.07 \\
12235 & 6100 & 4.40 & 1.60 & 0.35 & 0.11 \\
16160 & 5100 & 4.55 & 0.60 & $-$0.03 & 0.14 \\
16895 & 6500 & 4.70 & 1.70 & 0.07 & 0.08 \\
22484 & 6050 & 4.30 & 1.55 & $-$0.07 & 0.09 \\
26965 & 5300 & 4.55 & 0.70 & $-$0.26 & 0.08 \\
32147 & 5400 & 4.65 & 1.95 & 0.50 & 0.30 \\
48682 & 6200 & 4.60 & 1.30 & 0.13 & 0.13 \\
50281 & 5100 & 4.50 & 1.40 & 0.00 & 0.13 \\
76151 & 5700 & 4.35 & 0.65 & 0.10 & 0.05 \\
84737 & 5950 & 4.30 & 1.30 & 0.07 & 0.04 \\
126053 & 5650 & 4.40 & 0.65 & $-$0.44 & 0.04 \\
149661 & 5300 & 4.40 & 0.00 & 0.15 & 0.13 \\
157214 & 5650 & 4.35 & 0.60 & $-$0.44 & 0.06 \\
166620 & 5200 & 4.50 & 0.50 & $-$0.09 & 0.10 \\
170657 & 5200 & 4.55 & 0.60 & $-$0.11 & 0.08 \\
185144 & 5400 & 4.50 & 1.00 & $-$0.20 & 0.06 \\
186408 & 5780 & 4.35 & 0.85 & 0.09 & 0.04 \\
201091 & 5200 & 4.50 & 2.00 & $-$0.22 & 0.21 \\
219134 & 5100 & 4.40 & 0.90 & 0.10 & 0.13 \\
222368 & 6300 & 4.40 & 1.70 & $-$0.11 & 0.07 \\
\tableline
\end{tabular}
\end{center}
\end{table}

\subsection{Fe abundances}
\begin{figure}
\plotfiddle{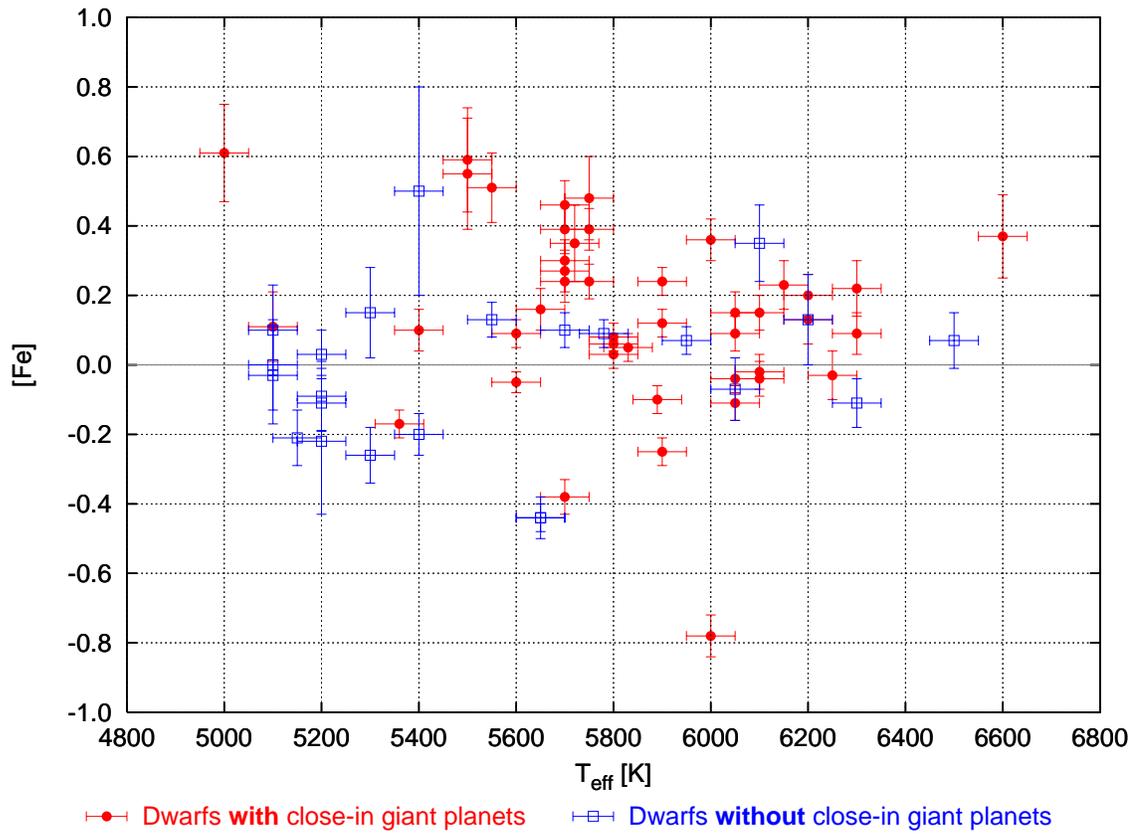}{11cm}{270}{60}{60}{-220}{350}
\caption{Fe-abundances (relative to the Sun) of CGP and no-CGP dwarfs.}
\label{results}
\end{figure}

The parameters and Fe abundances determined spectroscopically in this work are given in Tables~\ref{results1} and \ref{results2}.
Figure~\ref{results} shows the Fe abundances as a function of effective temperature for the two different groups of stars. For the calculation of the model atmospheres we used ODFs with solar abundances scaled according to the derived Fe abundances. We found this to be particularly important for overabundant stars, for which the abundances were underestimated when ODFs with solar abundances were used.

\section{Conclusions}
At this point of time we cannot draw any hard conclusions, because
\begin{itemize}
\item the sample sizes are limited;
\item the analysis is still incomplete;
\item the samples could be affected by selection bias.
\end{itemize}

\noindent However, Figure~\ref{results} indicates that
\begin{itemize}
\item in contrast to the no-CGP sample, there seems to be a lack of cool stars ($T_{\rm eff} \le$ 5400 K) in the CGP sample. Does this mean a bias in the sample selection?
\item there is a ``clump'' of metal-rich CGP stars at solar temperature. Is this a lack of metal poor giant planet hosts or a bias -- either from selection or from a lack of metal poor solar type stars (with or without planets)?
\item there seems to be {\em no} correlation between metallicity and planet hosting at hotter temperatures ($T_{\rm eff} \ge$ 6000 K).
\end{itemize}

\noindent Planned future work includes:
\begin{itemize}
\item Abundance analysis of a second comparison sample: Stars which have spectral types like those of the dwarfs with (CG) planets selected randomly from the Bright Star Catalog (and Supplement);
\item Abundance analysis of ``Very Strong Lined'' dwarfs (Eggen 1978);
\item Abundance determination for elements other than Fe, for all stars of the four samples.
\end{itemize}


\acknowledgments

This research has been supported by a grant from the National Science Foundation (NSF). 
Use was made of the Simbad database, operated at CDS, Strasbourg, France.

%
%

%

\end{document}